\documentclass[preprint,12pt]{elsarticle}
\usepackage{amsmath}
\usepackage{graphicx}  
\usepackage{epstopdf}   
\usepackage{lineno,hyperref}
\modulolinenumbers[5]
\usepackage{color}

\begin{document}

\begin{frontmatter}

\title{Effect of the entropy  on the shear viscosity of metallic glasses near the glass transition}

\author[VSPU]{A.S. Makarov}
\author[NWPU]{J.B. Cui}  
\author[NWPU]{J.C. Qiao} 
\author[VSPU]{G.V. Afonin}
\author[IFTT]{N.P. Kobelev}
\author[VSPU]{V.A. Khonik\corref{cor1}}
\cortext[cor1]{{Corresponding author}}
\ead{v.a.khonik@yandex.ru}

\address[VSPU] {Department of General Physics, State Pedagogical
University,  Lenin St. 86, Voronezh 394043, Russia}
\address[NWPU] {School of Mechanics and Civil Architecture, Northwestern Polytechnical University, Xi’an 710072, China}
\address[IFTT] {Institute of Solid State Physics, Russian Academy of Sciences, Chernogolovka, Moscow district 142432, Russia}

\begin{abstract}
We measured the shear viscosity of 14 metallic glasses differing with their mixing entropy $\Delta S_{mix}$. It is found that the viscosity at the glass transition temperature $T_g$ significantly increases with  $\Delta S_{mix}$. Using calorimetric data, we calculated the excess entropy  of all glasses $\Delta S$ with respect to their maternal crystalline states as a function of temperature. It is shown that the excess entropy $\Delta S$ both at room temperature and at $T_g$ \textit{decreases} with  $\Delta S_{mix}$. It is concluded that glasses with "high mixing entropy" $\Delta S_{mix}$ correspond to MGs with \textit{low} excess entropy $\Delta S$. The origin of  the increased shear viscosity at $T_g$ of glasses with high $\Delta S_{mix}$ is determined by their reduced excess entropy $\Delta S$.
\end{abstract}


\end{frontmatter}

The beginning of the 2000-s was characterized by the synthesis of new crystalline materials containing  five or more metallic elements each having the atomic percentage between 5 and 35 \%. These materials are usually characterized by the entropy of mixing (also called as the configurational entropy) defined as 

\begin{equation}
\Delta S_{mix}=-R\sum\limits_{i=1}^nc_ilnc_i, \label{Smix}
\end{equation}
where $R$ is the universal gas constant, $c_i$ is the molar fraction of the \textit{i}-th element and $n$ is the number of constituent elements \cite{YehAdvEngMater2004,CantorMaterSciEng2004}. These materials were named as high-entropy (HE) alloys. A few years later it was found that the melts of HE alloys can be solidified by quenching into  non-crystalline state forming thus HE metallic glasses (MGs) \cite{GaoJNCS2011,TakeuchiIntermetallics2011,WangJOM2014}. Since then, the number of works devoted to HE MGs has been increasing rapidly \cite{LuanJMaterSciTechnol2023}.

It has been shown that HE MGs have a supercooled liquid region of up to 65 K, low reduced glass transition temperature and display quite high glass-forming ability \cite{BizhanovaJALCOM2019,ChenJNCS2015,DingMaterLett2014,WadaMaterialia2019, OhashiJALCOM2022,WangJNCS2022}. Important features of HE MGs are related to their enhanced thermal stability \cite{CaoIntermetallics2018,DehkordiJNCS2022,HuoChinPhysLett2022,LuanNatComm2022}, higher activation energies  \cite{ChenIntermetallics2023},  lower atomic mobility   \cite{ChenIntermetallics2023}, sluggish dynamics of homogeneous flow \cite{ZhangScrMater2022},
 sluggish diffusion \cite{DuanPRL2022,JiangNatComm2021}, decreased dynamic and spatial heterogeneities  \cite{JiangNatComm2021}. It was noted that HE MGs have good  mechanical properties \cite{LuanJMaterSciTechnol2023,ChenJNCS2015,QiIntermetallics2015} and sometimes display superior magnetic characteristics \cite{QiIntermetallics2015,XuJNCS2018,ChenJALCOM2021}, superplastic behavior above the glass transition temperature $T_g$ \cite{DehkordiJNCS2022}, unique biomedical properties \cite{WangJOM2014} as well as excellent irradiation tolerance \cite{WangJNucLMater2019}. It is occasionally said that HE MGs combine the features of both metallic glasses and high entropy crystalline alloys  \cite{ChenJALCOM2021}. 
 
An important and interesting finding was presented by the authors \cite{LuanNatComm2022} who found that MGs' high entropy can enhance the degree of disorder and elevate it to a high-energy glassy state leading to a glass-to-glass phase transition upon heating.  This phase  transition is characterized by a heat release, which is even larger than that during following crystallization at higher temperatures and  leads to a significant improvement in the elastic modulus, hardness and thermal stability \cite{LuanNatComm2022}. 

Despite of intense investigations of HE MGs during the past decade \cite{LuanJMaterSciTechnol2023}, the progress in the understanding of the role of the high mixing entropy in their physical properties is clearly insufficient. On the one hand, the definition of the mixing entropy $S_{mix}$ given by Eq.(\ref{Smix}) accounts only for the geometrical disorder produced by different chemical elements in the melt but does not consider any kind of chemical interactions between constituent atoms, which should certainly affect the physical properties of solid glass. Meanwhile, it is reasonable to accept that these chemical interactions will give rise to differences in the properties of MGs even if they have the same $\Delta S_{mix}$. Besides that, a metallic glass has additional contribution to the entropy (both configurational and vibrational) due to structural defects and this contribution, moreover, can be dominating \cite{MakarovJPCM2021b,MakarovJETPLett2022}. From this viewpoint, one should consider the excess entropy of MGs with respect to the crystalline state, which by definition includes all components of the entropy for a particular metallic glass. We are unaware of any research in this direction with the exception of recent papers \cite{MakarovJPCM2021b,MakarovJETPLett2022}. In this work, we present, first, the effect of the mixing entropy on the shear viscosity of MGs at $T_g$ and, next, analyze this effect in terms of the excess entropy of these glasses.  

We studied 14 MGs produced by melt spinning as ribbons 15--30 $\mu$m thick (up to 50 $\mu$m in some cases)  and 1--2 mm wide, which were confirmed to be completely X-ray amorphous. The glass compositions are listed in Table 1. All glassy samples  were carefully studied by differential scanning calorimetry (DSC) using a Hitachi DSC 7020 instrument operating in high-purity (99.999\%) $N_2$ atmosphere at a rate of 3 K/min. 
Every glass was tested as follows: \textit{i}) initial sample 1 was heated up to the temperature of the complete crystallization $T_{cr}$ with empty reference DSC cell. This (crystallized) sample was next put to the reference DSC cell; \textit{ii}) a new sample 2 in the initial state was tested up to $T_{cr}$ and, finally, \textit{iii)} the same sample 2 was tested up to $T_{cr}$ (run 2). This protocol allows performing  measurements with the reference cell containing fully crystallized sample (prepared by the step \textit{i}) of approximately the same mass ($\approx$  70 mg) so that the measured heat flow $\Delta W$ constitutes the difference between the heat flow coming from glass $W_{gl}$ and its crystalline counterpart $W_{cr}$ and hereafter termed  as the differential heat flow, i.e. $\Delta W=W_{gl}-W_{cr}$.

\begin{table}[t]
\caption{\label{tab:table1} Metallic glasses under investigation and their mixing entropies $S_{mix}/R$ calculated with Eq.(\ref{Smix}). } 

\scriptsize
\begin{tabular}{p{3mm}|c|c|c|c|c|c|c|p{6mm}|c}
\hline
\hline
No & Composition (at.\%)&$\Delta S_{mix}/R$&No&Composition (at.\%)&$\Delta S_{mix}/R$ \\ 
\hline
\hline

 1 & Cu$_{50}$Zr$_{45}$Al$_5$ & 0.86 &8 &Pt$_{20}$Pd$_{20}$Cu$_{20}$Ni$_{20}$P$_{20}$ & 1.61\\

 2 & Cu$_{49}$Hf$_{42}$Al$_9$ & 0.93  &9&Zr$_{35}$Hf$_{13}$Al$_{11}$Ag$_8$Ni$_8$Cu$_{25}$&1.63 \\

3 & Zr$_{46}$Cu$_{45}$Al$_{7}$Ti$_2$    & 0.98&10&Zr$_{30}$Hf$_{25}$Al$_{20}$Co$_{10}$Ni$_{10}$Cu$_5$&1.64\\ 

4 & Zr$_{65}$Al$_{10}$Ni$_{10}$Cu$_{15}$ & 1.03 & 11 &Zr$_{40}$Hf$_{10}$Ti$_{4}$Y$_1$Al$_{10}$Cu$_{25}$Ni$_{7}$Co$_{2}$Fe$_{1}$&1.66\\

5 & Zr$_{50}$Cu$_{25}$Co$_{12.5}$Al$_{12.5}$ & 1.21 &   12&Zr$_{35}$Hf$_{17.5}$Ti$_{5.5}$Al$_{12.5}$Co$_{7.5}$Ni$_{12}$Cu$_{10}$&1.77\\

6 & Zr$_{52.5}$Ti$_5$Cu$_{17.9}$Ni$_{14.6}$Al$_{10}$ & 1.31 & 13&Ti$_{16.7}$Zr$_{16.7}$Hf$_{16.7}$Cu$_{16.7}$Ni$_{16.7}$Be$_{16.7}$&1.79\\

7 & Zr$_{31.6}$Hf$_{13.4}$Al$_{8.7}$Ag$_{8.4}$Cu$_{37.8}$&1.42&14 &Zr$_{16.7}$Hf$_{16.7}$Al$_{16.7}$Co$_{16.7}$Ni$_{16.7}$Cu$_{16.7}$&1.79\\
 
\hline
\hline

\end{tabular}
\end{table}

\begin{figure}[t]
\center{\includegraphics[scale=0.75]{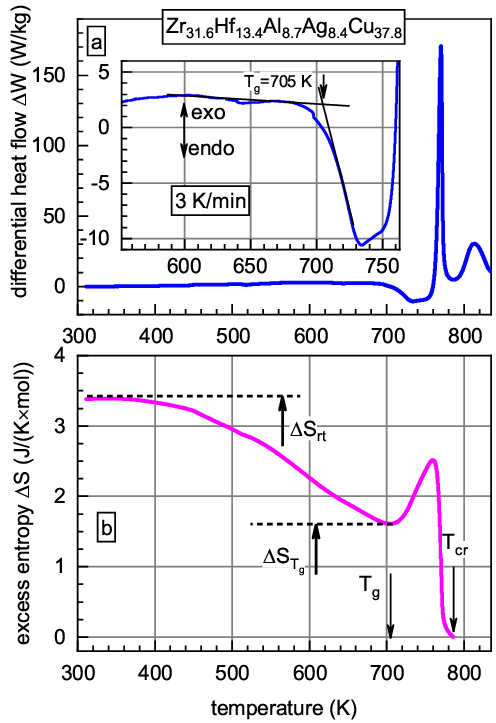}}
\caption[*]{\label{Fig1.eps} (a) Differential heat flow $\Delta W$ of glassy Zr$_{31.6}$Hf$_{13.4}$Al$_{8.7}$Ag$_{8.4}$Cu$_{37.8}$ (composition 7 in Table 1). The inset gives the portion of the thermogram near the glass transition region on an enlarged scale. The way of the determination of the glass transition temperature $T_g$ is exemplified. (b) Excess entropy $\Delta S$ calculated with Eq.(\ref{DeltaS}) using $\Delta W$-data shown in  panel (a). $\Delta S$-values at room temperature and  at $T_g$ are indicated. The temperature of the complete crystallization $T_{cr}$ is shown by the arrow.} 
\end{figure} 

The shear viscosity was determined using tensile creep measurements at the same rate of 3 K/min in a vacuum of about 0.01 Pa using a laboratory-made instrument \cite{MakarovJPCM2021a}. To compensate for significant thermal expansion of the instrument occurring upon heating, every viscosity curve was calculated using two creep runs taken on different initial samples loaded by two different tensile stresses, a low stress $\sigma_l\approx 10$ MPa and a high stress $\sigma_h\approx 120$ MPa (both set accurate to $\approx 10\%$). Changes in sample cross section $S$ occurring in the course of heating were taken into account upon viscosity calculations. For this purpose, the cross section was accepted as $S(T)=\frac{h_0d_0l_0}{l_0+\Delta l(T)}$, where $h_0$, $d_0$ (0.3 $mm$ to 0.6 $mm$), $l_0$ ($\approx 20$ $mm$) are the initial thickness, width and gauge length of a sample, respectively, and $\Delta {l}$ is the elongation. The shear viscosity was then calculated as $\eta=\frac{\sigma_{eff}}{3\;\dot{\varepsilon}_{eff}}=\frac{\sigma_h-\sigma_l}{3\;\frac{d}{dt}\left(\varepsilon_h-\varepsilon_l\right)}$, where $\varepsilon$ is the strain, the subscripts $h$ and $l$ correspond to the measurement runs under the high and low stresses, respectively.  The absolute error of elongation measurements  was about 0.01 $\mu m$, the elongation sampling frequency was $\approx 0.1$ Hz. The measurements were stopped above $T_g$ when the tensile strain $\varepsilon =\Delta l/l_0$ reached a value of 0.20.  The results given below were derived at an effective stresses $\sigma_{eff}=\sigma_h-\sigma_l$ of about 100-110 MPa that corresponds to linear Newtonian viscous flow. Some other experimental details are given in Ref.\cite{MakarovJPCM2021a}.

\begin{figure}[t]
\center{\includegraphics[scale=0.75]{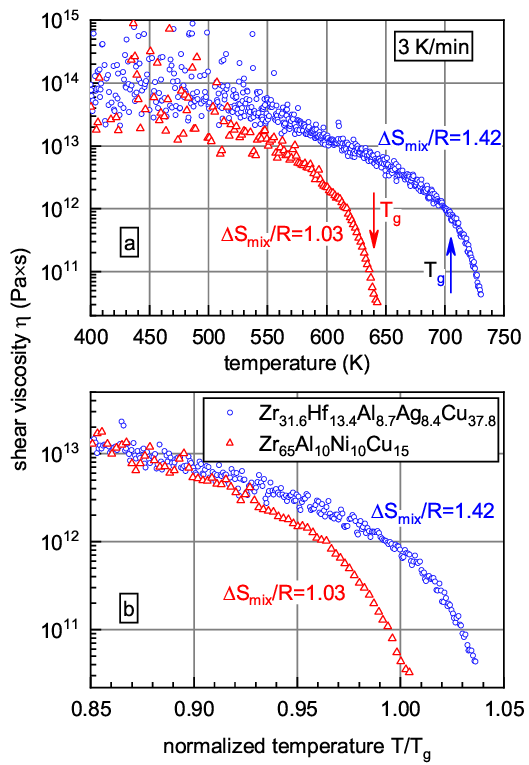}}
\caption[*]{\label{Fig2.eps} Logarithm of the shear viscosity $\eta$ of glassy Zr$_{31.6}$Hf$_{13.4}$Al$_{8.7}$Ag$_{8.4}$Cu$_{37.8}$ (composition 7) and Zr$_{65}$Al$_{10}$Ni$_{10}$Cu$_{15}$ (composition 4) as a function of temperature $T$ (a) and normalized temperature $T/T_{g}$ (b). It is seen that the viscosity at $T_g$ for the glass with bigger excess entropy $\Delta S_{sql}$ is significantly bigger.}  
\end{figure} 

The present investigation is essentially based on calorimetric data. As an example, Fig.\ref{Fig1.eps} gives a DSC trace of glassy Zr$_{31.6}$Hf$_{13.4}$Al$_{8.7}$Ag$_{8.4}$Cu$_{37.8}$, which shows a few features quite typical for MGs:  \textit{i}) exothermal heat flow below the glass transition, \textit{ii}) endothermal reaction in the supercooled liquid region above $T_g$ and \textit{iii}) strong exothermal crystallization-induced heat flow. These features are characteristic for all MGs under investigation.     

\begin{figure}[t]
\center{\includegraphics[scale=0.75]{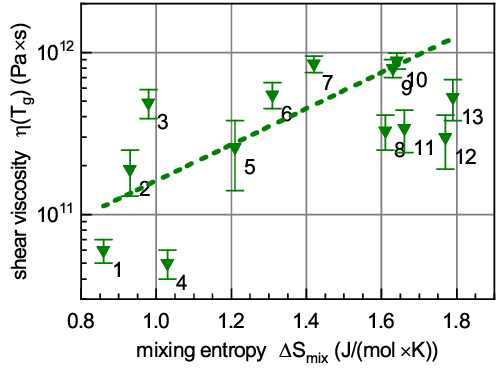}}
\caption[*]{\label{Fig3.eps}  Shear viscosity at the glass transition temperature as a function of the mixing entropy for MGs listed in Table 1.}  
\end{figure}     

Let us first assume that the shear viscosity $\eta$ in the $T_g$-region is determined by the mixing entropy $\Delta S_{mix}$. In this case, it is reasonable to compare $\eta$ of MGs with notably different $\Delta S_{mix}$. Figure \ref{Fig2.eps}(a) gives temperature dependence of  $log\;\eta$ for glassy Zr$_{31.6}$Hf$_{13.4}$Al$_{8.7}$Ag$_{8.4}$Cu$_{37.8}$ (composition 7) having the mixing entropy $\Delta S_{mix}/R=1.42$ ($T_g=705$ K) together with the measurements on glassy Zr$_{65}$Al$_{10}$Ni$_{10}$Cu$_{15}$  (composition 4) with a significantly smaller mixing entropy $\Delta S_{mix}/R=1.03$ ($T_g=640$ K). The corresponding glass transition temperatures  are indicated by the arrows. It is seen that $log\;\eta (T)$-dependences in general are quite similar and do not point to any peculiarities related to the mixing entropy. However, replotting the same data versus the normalized temperature $T/T_g$ given in Fig.\ref{Fig2.eps}(b) clearly shows that the viscosity of the first glass with bigger $\Delta S_{mix}$  in the vicinity of $T_g$ is bigger by more than an order of magnitude. 

This conclusion is further confirmed by the data in Fig.\ref{Fig3.eps}, which gives  $log\;\eta(T_g)$ as a function of the mixing entropy for all MGs under investigation. The numbers correspond to MGs' compositions according to Table 1 and the solid line gives a least square data fit (the same type of fitting is performed for other plots given below). It is clearly seen, despite of  data scatter, that $log \;\eta(T_g)$ \textit{increases} with $\Delta S_{mix}$. This fact seems to be strange since \textit{bigger} viscosity implies \textit{smaller }atomic mobility while the latter is reasonable to relate with bigger structural order and, hence, with smaller entropy of the structure. This  contradicts the data given in Figs \ref{Fig2.eps} and \ref{Fig3.eps}.  

We believe that the reason of this contradiction is associated with the mixing entropy, which reflects purely geometrical entropy change and does not  account for any chemical aspects of interatomic interactions in a complex system. Besides that, the mixing entropy does not consider any peculiarities of the glassy state since it does not change upon transition from crystalline  to non-crystalline  state.   It turns out that the mixing entropy $\Delta S_{mix}$ is not suitable as an independent variable for the understanding of MGs' property changes upon varying  chemical composition.  From this viewpoint, one should try another entropy-based parameter for this purpose. 

In this regard, we recently considered the excess entropy $\Delta S$ of glass with respect to the maternal crystalline state. In line with general thermodynamic definition of the entropy, this quantity is defined as \cite{MakarovJPCM2021b,MakarovJETPLett2022}

\begin{equation}
\Delta S(T)=\frac{1}{\dot{T}}\int_{T}^{T_{cr}} \frac{\Delta W(T)}{T}dT, \label{DeltaS}
\end{equation}
where $\Delta W$ is the differential heat flow specified above, $T_{cr}$ is the temperature of the complete crystallization and $\dot{T}$ is the heating rate. It should be emphasized that  if   temperature $T=T_{cr}$ then the integral (\ref{DeltaS}) turns to zero and, therefore, $\Delta S$ describes solely the excess entropy of glass with respect to the maternal crystalline state. It should also be pointed out that $\Delta S$ constitutes  the full excess entropy of glass, which includes both vibrational and configurational components.

\begin{figure}[h]
\center{\includegraphics[scale=0.75]{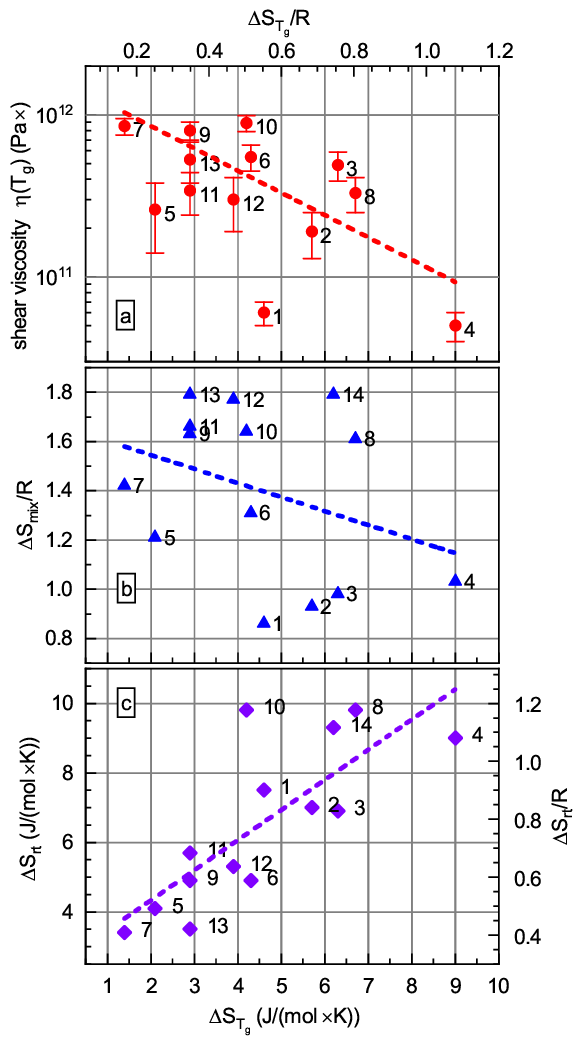}}
\caption[*]{\label{Fig4.eps} Shear viscosity $\eta (T_g)$ (a), mixing entropy $\Delta S_{mix}$ (b) and excess entropy at room temperature $\Delta S_{rt}$ (c) as a function of the excess entropy  $\Delta S_{T_{g}}$  at the glass transition temperature $T_g$ for indicated MGs.} 
\end{figure}

As an example, Fig.\ref{Fig1.eps}(b) gives temperature dependences of the excess entropy $\Delta S$ of glassy Zr$_{31.6}$Hf$_{13.4}$Al$_{8.7}$Ag$_{8.4}$Cu$_{37.8}$ calculated using Eq.(\ref{DeltaS}) using differential heat flow $\Delta W$ shown in panel (a) of this Figure. In the range from room temperature to $T_g$, the excess entropy decreases reflecting an increase of structural order during exothermal structural relaxation (see also the DSC thermogram in Fig.\ref{Fig1.eps}(a)) and reaches a minimum near $T_g$ indicated as $\Delta S_{T_g}$. Upon further heating, $\Delta S$ rapidly increases due to endothermal disordering in the supecooled liquid state, reaches a maximum  an then quickly falls down to zero because of crystallization. The changes of the excess entropy $\Delta S$ are induced by changes of structural order, as discussed in detail recently  \cite{MakarovScrMater2024}. It should be underlined that such $\Delta S(T)$-dependences  are characteristic of other MGs in this investigation, with the exception of HE glass 14 (Table 1). This glass in the initial state does not show any  $\Delta S$  maximum in the glass transition region before crystallization onset and its shear viscosity  near $T_g$ does not reveal any change of the slope, which is characteristic of other MGs (see Fig.\ref{Fig2.eps}).  This is  why this glass is not considered in Figs \ref{Fig3.eps} and \ref{Fig4.eps}(a).

On can now plot the shear viscosity at the glass transition as a function of the excess entropy. Since the viscosity is taken at $T=T_g$, it is reasonable to plot it versus the excess entropy $\Delta S_{T_g}$ also taken at $T_g$. Figure \ref{Fig4.eps}(a) shows that the viscosity $\eta (T_g)$ decreases with $\Delta S_{T_g}$ that is quite expectable since the bigger is the excess entropy, the bigger the degree of disorder is and the smaller the shear viscosity should be. The upper axis in Fig.\ref{Fig4.eps}(a) gives the excess entropy at $T_g$ in $R$-units. This leads to an interesting observation that the disordering at $T_g$ reflected by $\Delta S_{T_g}$  is significantly smaller than that characterized by the mixing entropy $S_{mix}$ (see Table 1). The same applies for the excess entropy at room temperature $\Delta S_{rt}$.

It is important to compare the excess entropy at the glass transition $\Delta S_{T_g}$ with the  mixing entropy $\Delta S_{mix}$.  This is done in Fig.\ref{Fig4.eps}(b), which shows that $\Delta S_{mix}$  \textit{decreases} with $\Delta S_{T_g}$. In a qualitative sence, the physical origin of this interdependence seems to be clear. The high mixing entropy means that the melt was strongly relaxed due to accomodative movements of five or more different atoms, which occupy many of potential wells in the structure. This relaxed melt  is then solidified into a glass, which remains relaxed to certain extent even in the supercooled liquid state (despite of  disordering in this state) providing a reduced value of $\Delta S_{T_g}$. 

Meanwhile, the excess entropy at room temperature $\Delta S_{rt}$ increases with the excess entropy at the glass transition $\Delta S_{sql}$ as verified by the data shown in Fig.\ref{Fig4.eps}(c). This means that the so-called "high entropy MGs" (i.e. those commonly accepted to have large $\Delta S_{mix}$) are actually the glasses with \textit{reduced excess entropy} $\Delta S_{rt}$ (or $\Delta S_{T_g})$. This is the main finding of the present work.

Since "high entropy MGs" have low excess entropy, they should display reduced ability for relaxation. In particular, it is clear that since glass   Zr$_{31.6}$Hf$_{13.4}$Al$_{8.7}$Ag$_{8.4}$Cu$_{37.8}$ (composition 7) has much smaller $\Delta S_{T_g}$ compared with that of Zr$_{65}$Al$_{10}$Ni$_{10}$Cu$_{15}$ (composition 4) (1.4 J/(mol$\times$K) and 9 J/(mol$\times$K), respectively, see Fig.\ref{Fig4.eps}(a)),  it should have bigger viscosity at $T_g$, as indeed observed.  We, therefore, predict that relaxation-resistant MGs should be found amongst MGs with high mixing entropy or, equivalently, with low excess entropy (either at room temperature or at $T_g$). Low atomic mobility, sluggish dynamics of homogeneous flow and sluggish diffusion of HE MGs with large $\Delta S_{mix}$ described in the literature and mentioned above support this prediction.
 
In conclusion, we performed measurements of the shear viscosity $\eta$ of 14 MGs with different mixing entropies $\Delta S_{mix}$. It is found that $log\; \eta (T=T_g)$ increases with $\Delta S_{mix}$. This physically unclear dependence indicates that  $\Delta S_{mix}$ cannot be considered as a proper physical parameter for the understanding of glass properties upon varying chemical composition. Instead, as a suitable parameter for this understanding we used temperature-dependent excess entropy of glass with respect to the maternal crystal $\Delta S$ defined by Eq.(\ref{DeltaS}). It is shown that $log\; \eta (T_g)$ \textit{decreases} with $\Delta S$ (determined either at room temperature or at $T_g$), as expected. At that, the excess entropy $\Delta S$ decreases with the mixing entropy $\Delta S_{mix}$. This means that "high entropy metallic glasses" (commonly understood as MGs having high $\Delta S_{mix}$) actually constitute glasses with low excess entropy $\Delta S$ and, consequently, with reduced relaxation ability. In particular, this is manifested in a decrease of the shear viscosity at $ T_g$ with increasing $\Delta S$.

This work was supported by the Russian Science Foundation under the grant No 23-12-00162.

\vspace{6pt}




\end{document}